# Vertical transport and tunnelling in rare-earth nitride heterostructures

Jackson D. Miller, Felicia H. Ullstad, H. Joe Trodahl, Ben. J. Ruck and Franck Natali

The MacDiarmid Institute for Advanced Materials and Nanotechnology, School of Chemical and Physical Sciences, Victoria University of Wellington, P.O. Box 600, Wellington 6140, New Zealand

E-mail: franck.natali@vuw.ac.nz

**Abstract**

We report an investigation of the ferromagnetic semiconductor rare earth nitrides (RENs) for their potential for cryogenic-temperature electronics and spintronics application. We have indentified ohmic contacts suitable for the device structures that demand electron transport through interface layers, and grown REN/insulator/REN heterostructures that display tunnelling characteristics, an enormous 400% tunneling magnetoresistance and a hysteresis promising their exploitation in non-volatile magnetic random access memory.

**1. Introduction**

It is now well-established that the rare earth nitrides (RENs) form a series of intrinsic ferromagnetic semiconductors suitable for full exploitation of the coupled magnetic/electronic technology, so-called spintronics [1]. They display their magnetic properties only at low temperatures (<70 K) and while it prevents room temperature applications, they are materials of choice on the more demanding route to devices for cryogenic electronics [2]. Their complementary properties, in particular the varied set of magnetic properties across the series, provide huge scope for use in memory elements such as non-volatile magnetic random access memory (MRAM) to support superconducting and quantum computing [3-10]. In addition, a strong impetus to select RENs over ferromagnetic metals or diluted magnetic semiconductors (DMS) derives from the ability to dope the materials without losing ferromagnetic spin alignment [11,12]. This offers potential integration and impedance matching with the materials and processing for superconducting electronics. However, only little has been reported concerning the study of vertical transport, the so-called Current Perpendicular to Plane (CPP) geometry, [13-16] with the literature focussing mostly on electrical transport properties in planar structures for fundamental studies [17-21]. Since devices will mostly require electronic transport across material interfaces, there is an urgent need to provide a general guideline for the design and analysis of vertical tunnelling devices and heterostructures if the RENs are to be technologically exploited. Remaining hurdles like ensuring ohmic contacts, developing lithographic procedures and reducing device size towards a practical 1μm$^2$ target have to be overcome. Progress towards all of these goals are presented and discussed in this paper. In particular, we show evidence for ohmic contact to GdN, the prototypical REN, in CPP device geometries, and optically defined device areas down to $10 \times 10$ μm$^2$ without oxidation of the GdN layer. Finally a clear hysteresis at low magnetic field is oberved in magnetic tunnelling junctions, a promising advance towards the development of REN-based memory element structures.

**2. Methods**

In this study, heterostructures consisting of metal/GdN/metal thin layers and magnetic tunnelling devices consisting of metal/REN/insulator/REN/metal thin layers have been fabricated using a multi-step approach on 300 nm thick SiO$_2$ layers thermally grown on silicon (100) wafers. The metal/GdN/metal structures were used as a model system to determine the nature of the contact, ohmic or Schottky, between a range of metallic layers, here Al, Au and Gd, and the GdN semiconductor layer. Due to their wide band gap and their chemical and structural compatibility with the RENs, GaN or AlN are used as insulating layers to form tunnel barriers. We have chosen GdN and SmN among the fourteen RENs due to their contrasting magnetic properties [1,16].

RENs have been recognized for decades to react rapidly in air, requiring careful capping. Within this scenario REN-based CPP structures require special attention to avoid oxidation from open edges exposed to air that then slowly react and degrade the device performances. Figure 1 schematically shows the fabrication steps we developed to avoid the problem. Special attention was given, based on a so-called *caisson*-like structure [Figure 1(a)], both to overcome the oxidation problem and to reduce the size of devices from the 100 × 200 µm that we have previously achieved [16]. Our tunnelling devices are defined by small apertures (*caisson*) in a GaN layer that expose a segment of an underlying bottom metal contact. The stack then consists of the bottom metal contact strip (i.e. Al, Au or Gd), a 40 nm thick GaN insulator (with square *caissons* of 10, 20, 30 or 40 µm), the tunnelling stack (i.e. GdN/GaN/GdN or GdN/AlN/SmN) and top metal contact (i.e. Al, Gd); see Figure 1(b). The REN/top metal contact stack (step 3) fills the active area, allowing direct contact between the metallic bottom contact (step 1) and the stack. The GaN layer that defines the *caisson* (step 2) is grown thinner than the stack to ensure layer continuity across the active area edge. The surface profile across a GaN *caisson* region is shown in Figure 1(c). The two steep slopes at 600 and 640 µm are the *caisson* edges, so that the 35 nm dip in the middle shows the active area window depression that defines the device. Up to 20 lithographically defined tunnelling devices were prepared on a single chip, thus providing numerous devices of various sizes in a single preparation.

The REN (SmN, GdN), insulating (GaN, AlN) and top metal contact (Al, Gd) layers were grown at ambient temperature in a molecular beam epitaxy system with a base pressure <10$^{-9}$ Torr. The ambient growth temperature allows GdN and SmN layers to be smooth and strongly (111) textured. Au was used only as a bottom contact layer and was grown in a different physical vapor evaporator. Further details regarding the growth conditions (deposition rate, nitrogen precursor, etc) for the REN and insulating layers can be found elsewhere [16,18].

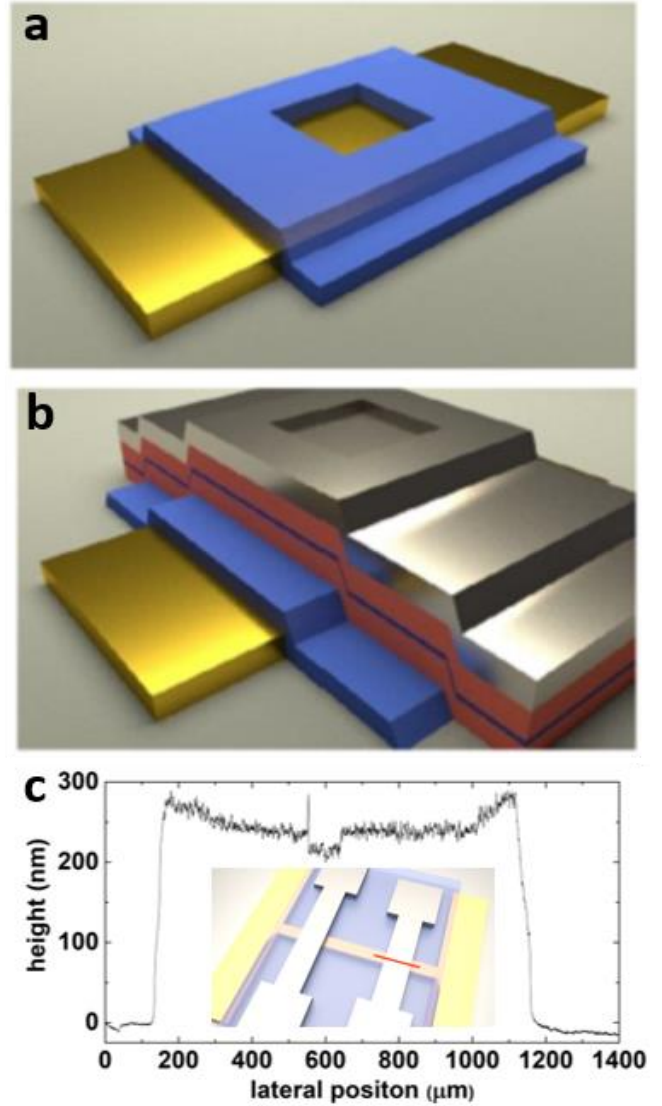

Figure 1. Schematic representation of the fabrication steps; (a) GaN *caisson-like* structure (blue) grown over the metal bottom contact (yellow) and (b) entire structure with two REN layers (red) separated by the tunnel barrier (black) and finished with the top contact (grey). (c) Surface profile of the stack in a metal/REN/insulator/REN/metal structure along the red line shown in the inset. The dip in the middle of the surface profile is where the stack is grown on top of the active area *caisson*.

The current/voltage (I/V) characteristics of the structures were measured between the crossed metallic strips, each offering separate voltage and current leads. The (I/V) measurements of the devices were performed between ambient temperature and 2 K in both a simple closed-cycle cryostat and a Quantum Design Physical Property Measurement System (PPMS) with the facility to apply magnetic fields to 9 T. Care was taken to ensure that the metal strips were sufficiently conductive to form virtual equipotentials, so that in turn the current densities through the devices were uniform, as was established by noting that the net current in most devices was proportional to the *caisson* area. To ease comparison of structures with different areas we report the results below in terms of the current density (J) rather than the measured net current.

## 3. Results

### 3.1. Metal/GdN/Metal structures

In view of the propensity for Schottky barrier formation across metal-semiconductor interfaces it is important to establish that the measured impedances of the tunnelling structures are dominated by the very thin tunnelling barrier layer, rather than the metal/GdN interfaces. We have thus investigated several structures in which the barrier layer was omitted to determine the resistance of metal/GdN/metal stacks. For technical convenience we have invested contacts comprising Al, Au and Gd.

Turning first to Al contacts, there is no sign of the non-linearity commonly associated with Schottky barriers, and the linear resistance observed yields a resistivity of the 100 nm-thick GdN layer of ~3 $\Omega$.cm at 4 K (not shown). That value is within the range that is commonly measured for in plane GdN layers, characterstic of a carrier density of order $10^{19}$ cm$^{-3}$ [12,21]. The impedance ratio (voltage/current density; V/J) provides a direct estimate of the impact the GdN resistance has on the tunnelling studies below. The data give V/J ~ $10^{-4}$ $\Omega$.cm$^2$ at low temperatures, which will be seen below to be six orders of magnitude smaller than the impedance ratio found in the tunnelling structures.

We have previously reported that Au forms an ohmic contact in measurements of in-plane transport [22], a geometry which is less sensitive to Schottky barrier effects due to its larger form factor (length/area), as we have also cofirmed in the present geometry [16]. Clearly both Al and Au form ohmic contacts to GdN.

The assessment of Gd metal as a possible contact material is critical as the exchange interaction with GdN or SmN across the material interface provides the pinning of the GdN magnetisation allowing observation of switching behaviour in a magnetic tunnnelling junction [16]. Similarly to the case of Al and Au, a Gd/GdN/Gd device structure was studied. As can be seen in Figure 2, the J/V plot is linear at ambient temperature but develops a modest non-linearity at low temperature. The impedance ratio (V/J) ~ $3 \times 10^{-3}$ $\Omega$.cm$^2$ at 4 K is a factor of $10^2$ larger than Au/GdN/Al structures. The larger resistance suggests the formation of Schottky barriers at the Gd/GdN interface which substantially deplete the carrier concentration within GdN. The non-linearity of the lower temperature is also an indication of a tunnelling-mode transport across the interfaces of the structure. While the conduction mode is not ohmic, the specific resistance remains three orders of magnitude smaller than the minimum impedance ratio across the tunnelling barrier structure we will see below, thus permitting the use of metallic Gd as a contact material.

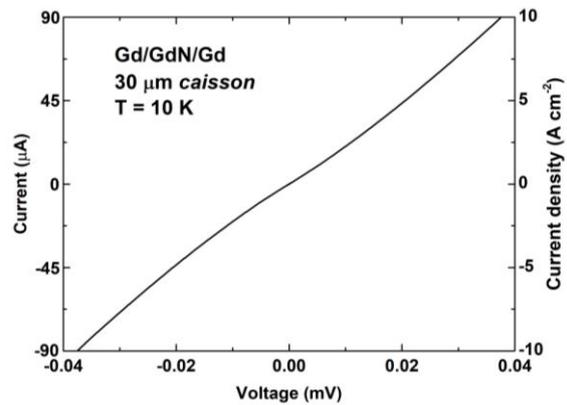

Figure 2. I/V and J/V characteristics of a Gd/GdN/Gd structure at 10 K showing only weak nonlinearity.

### 3.1. Tunnelling through engineered barriers

In Figure 3 we show the strongly nonlinear response of a Au/GdN/GaN/SmN/Al structure with target layer thicknesses of 100 nm GdN and SmN and 3 nm GaN. Even at the largest voltage, 0.9 V, the impedance ratio of order $6\times10^{-2}$ $\Omega$.cm$^2$ is still almost three orders of magnitude larger than that from the device without an engineered barrier. The data of Figure 3 have thus been fitted to the Simmons model [23,24] to yield estimates of ~1 eV and ~2 nm, respectively, for the GaN barrier height and width. These are within the range expected for the 3.3 eV band gap of GaN and the designed 2.5 nm barrier thickness. There is an asymmetry as expected with the GdN/SmN asymmetry of the source/drain for the tunnelling current.

Figure 3. I/V and J/V characteristics of a 60×60 (μm)$^2$

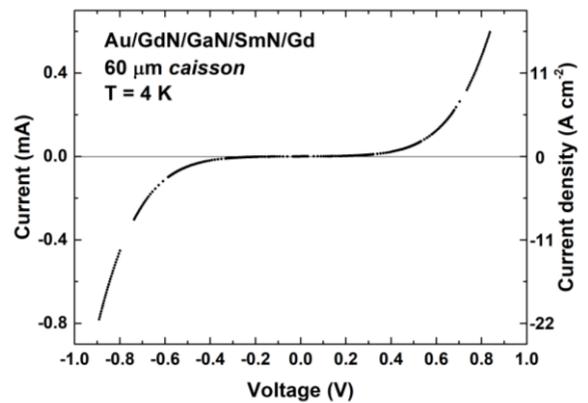

Au/GdN/GaN/SmN/Al structure at 4 K.

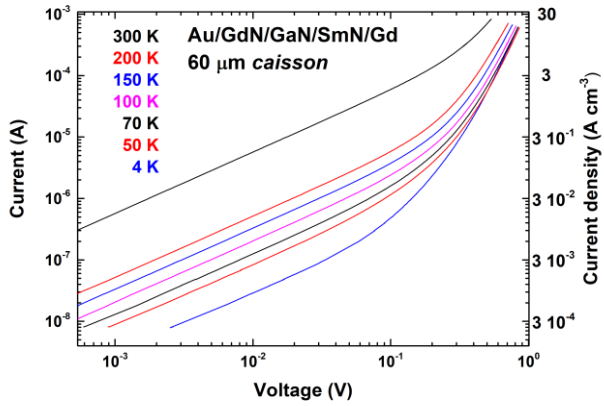

Figure 4. Temperature-dependent I-V and J/V characteristics of the device in Figure 3. These data were collected under a positive applied voltage, with current flowing from GdN to SmN.

The temperature dependant I/V response of the forward-biased device is shown at temperatures up to 300 K in Figure 4. The response is again seen to be strongly nonlinear, yielding Simmons-model estimates of ~1 eV and ~2 nm for the barrier height and width that are at most only weakly temperature dependent. The data show a linear behaviour at low voltage, where the conductance increases monotonically with temperature. In contrast the nonlinear conduction at larger applied voltage shows a complex temperature dependence, signalling clearly the shifting band edges across the Curie temperatures of GdN ($T_C$ ~ 70 K) and SmN ($T_C$ ~ 30 K) [1].

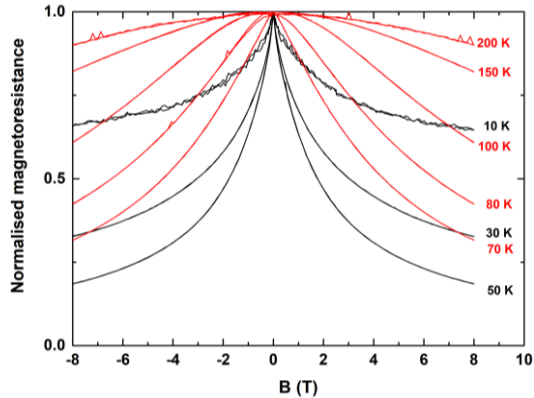

Figure 5. Tunnelling magnetoresistance for the device of Figures 3 and 4 at different temperatures, 1μA excitation.

The influence of spin alignment in the GdN and SmN layers is seen in the magnetic-field control of resistance of Figure 5. Note that the conventional characterisation of tunneling magnetoresistance is quoted as

$$\text{TMR} = [R(B) - R(max)]/R(min),$$

where $R(max)$ and $R(min)$ are the maximum and minimum resistance of the device, close to 0 Tesla and 8 Tesla in the present case. Thus the five-fold reduced resistance at 50 K corresponds to a TMR of 400%.

Of more direct interest for memory element devices is to establish the tunneling resistance in the very low-field region, where hysteresis exerts an influence. Thus in Figure 6 the hysteresis is displayed at 2 K, plotted as

$$\delta R = [R_{up,down}(B) - R_{ave}(B)] \equiv \pm[R_{up}(B) - R_{down}(B)]/2,$$

where $R_{up}$ and $R_{down}$ are the resistances under increasing and decreasing magnetic fields, respectively. It is expected that the relative spin alignments on the two sides of the barrier influence the measured resistance, and indeed here there are evident transitions near the coercive fields of GdN (~0.02 T) and SmN (>2 T) [1]. The switching shown in Figure 6 persists up to 30 K, the $T_C$ of SmN, demonstrating that the hysteresis behaviour requires both the GdN and SmN layers to be in their ferromagnetic states.

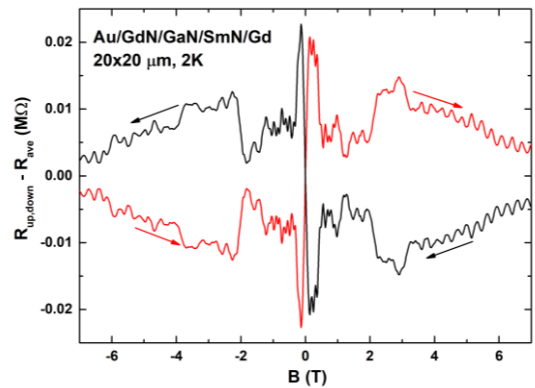

Figure 6. Magnification of low field hysteretic behaviour for an Au/GdN/AlN/SmN/Al structure.

## 4. Summary

We have reported growth of thin films of ferromagnetic semiconducting rare-earth nitrides in layered structures, designed for the investigation of the electronic reponse in current perpendicular to plane (CPP) configurations. Special care was taken to ensure that the devices were small enough to show excellent CPP impedance.

The work identified three contact metals (Au, Al, Gd) and investigated their susceptability to Schottky-barrier problems. Contacts comprising Au and Al show ohmic behaviour, while there is evidence of a depressed carrier concentration in GdN at an interface with Gd.

Identification of suitable ohmic-contact metals facilitated investigation of structures in which insulating tunnel barriers, 1-2 nm of AlN or GaN, were introduced between two REN layers, thus forming a magnetic tunnel junction. The structures showed clear negative tunnelling magnetoresistance as large as 400% (i.e. a contrast factor of 5) between fields of 0 and 9 T.

Finally the work has demonstrated clear tunnelling-resistance hysteresis, opening the door to devloping tunneling magnetoresistance memory for cryogenic memory banks.

## Acknowledgements

We acknowledge funding from the Marsden Fund (Grant No.13-VUW-1309), the New Zealand Endeavor fund (Grant No. RTVU1810) and the MacDiarmid Institute for Advanced Materials and Nanotechnology, funded by the New Zealand Centres of Research Excellence Fund. Jackson Miller thanks the School of Chemical and Physical Sciences for financial support.